\documentclass[aps, prstab, reprint, showpacs, amsmath, amssymb]{revtex4-1}

\usepackage{amsmath,amscd}
\usepackage{bm}
\usepackage[english]{babel}
\usepackage{graphicx}
\usepackage{soul}
\usepackage[usenames]{color}

\begin{document}
\title{ Radiation of a Charge in Dielectric Concentrator \\ for Cherenkov Radiation: Off-Axis Charge Motion }

\author{Sergey N. Galyamin}
\email{s.galyamin@spbu.ru}
\affiliation{Saint Petersburg State University, 7/9 Universitetskaya nab., St. Petersburg, 199034 Russia}

\author{Viktor V. Vorobev}
\affiliation{Saint Petersburg State University, 7/9 Universitetskaya nab., St. Petersburg, 199034 Russia}

\author{Andrey V. Tyukhtin}
\affiliation{Saint Petersburg State University, 7/9 Universitetskaya nab., St. Petersburg, 199034 Russia}

\date{\today}

\begin{abstract}
A theoretical investigation of radiation field produced by a charge moving through the dielectric concentrator for Cherenkov radiation is performed for the general case where a charge trajectory is shifted with respect to the target axis.
The idea of dielectric target with specific profile of the outer surface was presented and investigated in our previous papers for the symmetric case.
Here we show how non-symmetric field components generated in the bulk of target affect field distribution near the focus where strong concentration of the energy occurs.
Possible applications of this target are discussed.
\end{abstract}

\maketitle

\section{Introduction\label{sec:intro}}

Uniform movement of a charged particle in dielectric medium with velocity exceeding the speed of light in this medium is accompanied by Cherenkov radiation (CR) discovered in 1937~%
\cite{Ch37}
and theoretically interpreted the same year~%
\cite{TF37, Jb}.
Since that time this effect is widely used in various areas of physics~%
\cite{Jb, Zrb}
including relatively novel medical applications based on CR in biological tissues~%
\cite{Spin13}.

Today several modern trends based on beam-dielectric interaction exist in accelerator physics.
First, one should mention dielectric wakefield acceleration technique which is now operating with Terahertz (THz) wakefields~%
\cite{Ant12, JingAntipov18}
and has demonstrated Gigavolt per meter fields~%
\cite{OShea16}.
Dielectric-lined waveguides (or metallized capillaries) of various transverse cross-section, i.e. closed structures with dielectric, are utilized within the mentioned scheme.
Second, similar waveguide structures loaded with dielectric are considered nowadays as prospective candidates for contemporary beam-driven sources of THz radiation~%
\cite{Ant13, GTAB14, WangAntipov17}.
As was shown, with a proper electron beam and dielectric structure the THz radiation of an extraordinary peak power can be potentially obtained~%
\cite{OShea16}.

On the other hand, various opened dielectric structures (without metallization at the outer surface) are extensively studied nowadays for the development of both beam-driven radiation sources (mainly in THz range) and non-invasive systems of bunch diagnostics (including large-scale facilities such as LHC).
Diagnostics systems based on CR from prolonged dielectric targets possess several advantages compared to traditional schemes based on transition or diffraction radiation.
For example, a prismatic dielectric target was used in a series of experiments~%
\cite{Pot10, Kieffer18}
where prominent possibilities to use this scheme for beam position monitoring for high-energy electron and hadron beams were demonstrated.
Moreover, both mentioned prismatic target and hollow conical target can be a powerful source of radiation, including THz range~%
\cite{Takahashi00, Bergamproc17, Sei17}.

Although the theory of CR in infinite or semi-infinite dielectric media and dielectric-lined waveguide structures is well developed~%
\cite{B57, B62, Ab},
rigorous theoretical explanation of the radiation processes emerging during the interaction of charged particle with dielectric object of finite size (such as prismatic or hollow conical targets mentioned above) is extremely complicated.
Therefore, various approximate methods for calculation of radiation field are developed
\cite{Tishc04, Kar11, SKA15, PotGogSukh17}.
One of them is our original combined approach based on exact solution of certain ``etalon'' problem and consequent taking into account the outer boundary using ray-optics laws and Stratton-Chu formulas~%
\cite{BTG13, GT14, BGT15, TGV19, TVGB19}.
It is worth noting that recent papers~%
\cite{TGV19, TVGB19}
dealt with conical and prismatic objects which are of essential interest today due to the aforementioned applications.
The main advantage of this combined approach is its asymptotic accuracy with respect to the ratio between the wavelength and the size of the target.
Equally important is the fact that this method has been recently approved by numerical simulations in COMSOL Multiphysics~%
\cite{GTV18}.

In short, the mentioned approach consists of two steps.
At the first step we take into account only the single interface which is mainly interacts with a moving particle.
For example, this interface is the flat surface closest to the charge trajectory in the case of prismatic target or the cylindrical surface of the channel in the case of hollow conical target.
All other target surfaces are excluded from consideration at this step and we arrive to the corresponding ``etalon'' problem.
Known rigorous solution of this problem describes CR in the bulk of the target.
At the second step, we return the outer surface which is illuminated by CR. 
Mainly, radiation escapes the target through this surface (so called ``aperture''). 
We consider interaction of Cherenkov waves with this ``aperture'' using ray-optics laws.
In some cases, ray-optics consideration is suitable for description of the radiated field outside the target.
However, if we are interested in the radiated field either in the Fraunhofer zone or in the vicinity of focuses and caustics (as in the present paper) where ray-optics formalism fails, Stratton-Chu formulas can be utilized.
Input values for these formulas are tangential components of electric and magnetic fields at the outer side of the ``aperture'' which are calculated using Snell's law and Fresnel coefficients.

Concerning the topic of the present paper, the discussed approach was utilized to find the outer profile of the axisymmetric dielectric target concentrating the majority of generated CR in a small vicinity of a predetermined focus point without any additional lenses or mirrors~%
\cite{GT14}.
This target was called ``dielectric concentrator for CR'' and was investigated in details for symmetric case (charge movement along the target symmetry axis), including the study of sensitivity with respect to the charge velocity~%
\cite{GTV17, GTV18}.
However, for the practical realization it is also important to analyze the influence of the trajectory deviation off the symmetry axis on the radiation characteristics, which is the main goal of the paper.

The paper is organized as follows.
In Sec.~%
\ref{sec:problemstat}
we formulate the problem for the EM radiation from a shifted moving charge in the concentrator for CR, while for reader's convenience we clarify the determination of the target shape in Appendix~%
\ref{app:concdetails}%
.
The application of Stratton-Chu formulas is presented in Sec.~%
\ref{sec:SChu}
while the rigorous solution of ``etalon'' problem is given in Appendix~%
\ref{app:etalon}%
.
Section~%
\ref{sec:num}
contains numerical results while Sec.~%
\ref{sec:discuss}
discusses possible applications.
Section~%
\ref{sec:concl}
finishes the paper.

%
\begin{figure*}[t]
\centering
\includegraphics[width=0.85\linewidth]{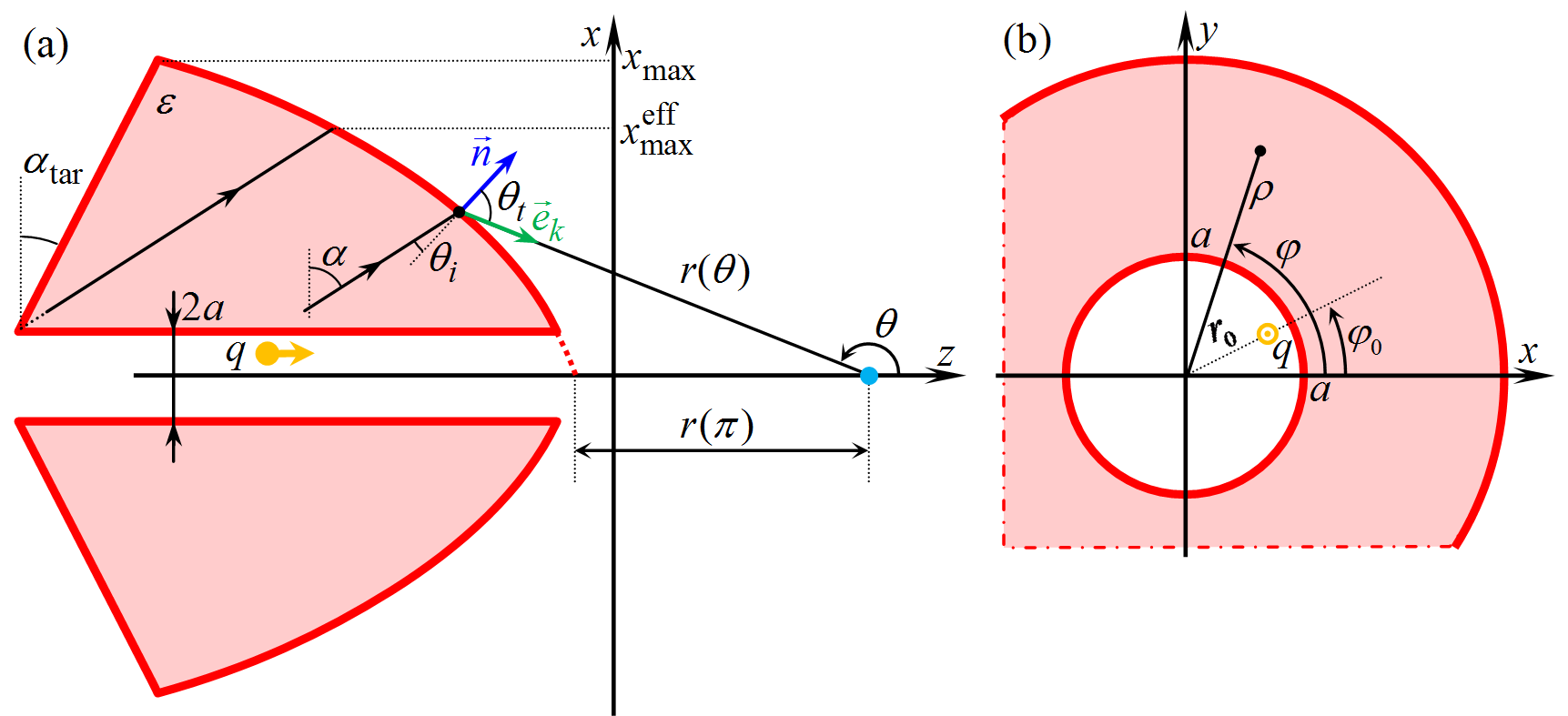}
\caption{\label{fig:geom}%
Geometry of the problem.
(a)
$ (zx) $%
-cut of the dielectric concentrator for Cherenkov radiation: the axisymmetric dielectric target with the specific outer profile (determined by the function
$ r( \theta ) $%
) and the inner channel of the radius
$ a $
where the point charge
$ q $
moves along the straight trajectory shifted with respect to the
$ z $%
-axis. 
Depicted parameters are discussed in the text.
(b) $ (xy) $%
-cut of the target (the channel radius is enlarged for convenience) and the position of the charge shifted trajectory.
}
\end{figure*}

\section{Problem formulation\label{sec:problemstat}}

Figure~%
\ref{fig:geom}
shows the geometry of the problem under investigation.
Note that along with Cartesian frame
$ ( x, y, z ) $,
corresponding cylindrical frame
$ ( \rho, \varphi, z ) $
is introduced.
A point charge
$ q $
moves with a constant velocity
$ \upsilon = \beta c $
along straight trajectory inside the channel in the axisymmetric dielectric target with permittivity
$ \varepsilon $
and permeability
$ \mu = 1 $.
Position of charge trajectory is determined by
$ r_0 $
and
$ \varphi_0 $%
, see Fig.~%
\ref{fig:geom}~%
(b).

Cylindrical coordinates
$ \rho = \rho_0 $,
$ z = z_0 $
of the outer profile of the target are determined for
$ r_0 = 0 $
as follows.
The point charge moving along the axis of the structure (symmetric case) with the velocity
$ \upsilon = \beta c $
generates Cherenkov radiation inside the target with the corresponding rays being directed at the angle
$ \alpha = {\rm arcsin }[ 1 / ( \sqrt{ \varepsilon } \beta ) ] $
with respect to vertical direction, these rays are shown in Fig.~%
\ref{fig:geom}~%
(a).
It is supposed here that Cherenkov condition is fulfilled,
$ \sqrt{ \varepsilon } \beta > 1 $.
The surface
$ \rho_0 $,
$ z_0 $
is determined by the rule that all the refracted rays converge exactly into the point
$ x = y = 0 $,
$ z = z_f $.
Details can be found in~%
\cite{GT14, GT17, GTV18}
and in the Appendix~%
\ref{app:concdetails}%
, in the issue we obtain the following
$ \varphi $%
-independent surface:
\begin{equation}
\label{eq:concsurf}
\begin{aligned}
\rho_0 ( \theta ) &= r( \theta ) \sin ( \theta ), \\
z_0 ( \theta ) &= z_f + r( \theta ) \cos ( \theta ),
\end{aligned}
\end{equation}
where
\begin{equation}
\label{eq:rTheta}
r( \theta )
=
f ( 1 - \sqrt{ \varepsilon } ) \left[ 1 + \sqrt{ \varepsilon } \sin ( \alpha + \theta ) \right]^{ -1 }.
\end{equation}
Here parameter
$ f $
can be called the ``focal'' parameter since the distance between the focal point and  the ``nose'' of the target
$ r( \pi ) $
is proportional to
$ f $%
, i.e.
\begin{equation}
\label{eq:rpi}
r( \pi )
=
\frac{ f \beta ( \sqrt{ \varepsilon } - 1 ) }{ 1 - \beta }.
\end{equation}
Therefore, the larger
$ f $
the larger distance between the focus and the target.
The maximum transverse size of the target
$ x_{ \max } $
determines the minimum angle
$ \theta_{ \min } $,
the maximum angle
$ \theta_{ \max } $
is determined by the channel radius
$ a $.

For the symmetric case (%
$ r_0 = 0 $%
) the radiation properties have been investigated in details, including dependence of the field components behavior on deviation of actual charge velocity from the ``designed'' velocity (sensitivity)~%
\cite{GTV17, GTV18}.
Moreover, the comparison between COMSOL Multiphysics simulations and the developed theory has been performed~%
\cite{GTV18}
and very well agreement has been observed.
Below we will investigate the influence of the offset
$ r_0 $
on the radiation characteristics of the concentrator.

\section{Stratton-Chu formalism\label{sec:SChu}}

According to our method we utilize the Stratton-Chu formulas~%
\cite{SChu39, Fradb}
to calculate Cherenkov radiation exiting the target.
Recall that these integral formulas give exact result if tangential electric and magnetic fields are determined exactly at the surface of the integration (the aperture
$ S _a $%
).
In this paper we use the form of these formulas from~%
\cite{Fradb}
(see also our papers~%
\cite{GT17, TGV19, TVGB19}%
) with the outer surface of the target~%
\eqref{eq:concsurf}
as the aperture:
\begin{equation}
\label{eq:SChu}
\begin{aligned}
&4 \pi
\vec{ E }_{ \omega }
=
\int\nolimits_{ S_a }
\left\{
\vphantom{ \frac{ i }{ k_0 } }
i k_0
\left[
\vec{ n }, \, \vec{ H }_{ \omega }^a
\right]
\psi
+
\right. \\
&+
\left.
\frac{ i }{ k_0 }
\left(
\left[
\vec{ n }, \, \vec{ H }_{ \omega }^a
\right], \, \vec{ \nabla }
\right)
\!
\vec{ \nabla }
\psi
+
\left[
\left[
\vec{ E }_{ \omega }^a, \, \vec{ n }
\right], \, \vec{ \nabla } \psi
\right]
\right\}
d \Sigma,
\end{aligned}
\end{equation}
where
$ \psi $
is a Green function,
\begin{equation}
\label{eq:Green}
\begin{aligned}
\psi
&= \left. \exp \left( i k_0 \tilde{ R } \right) \right/ \tilde{ R }, \\
\tilde{ R }
&=
\sqrt{ ( x - x_0 )^2 + ( y - y_0 )^2 + ( z - z_0 )^2 },
\end{aligned}
\end{equation}
$ d \Sigma $
is a surface element of
$ S_a $.
As follows from Eq.~%
\eqref{eq:SChu},
electromagnetic (EM) field outside the target is determined by the tangential electric and magnetic fields at the aperture
$ S_a $.
We utilize the following parametrization of the Cartesian coordinates of the aperture via angles
$ \theta $
and
$ \varphi $:
\begin{equation}
\label{eq:x0y0param}
x_0( \theta, \varphi ) = \rho_0( \theta ) \cos \varphi,
\quad
y_0( \theta, \varphi ) = \rho_0( \theta ) \sin \varphi,
\end{equation}
while
$ z_0 ( \theta, \varphi ) $
is given by~%
\eqref{eq:concsurf}
together with
$ \rho_0 ( \theta ) $.
In order to calculate the parameters of the surface it is convenient to use the tensor formalism by V.A.~Fock~%
\cite{Fockb, GTV17}
and determine the metric tensor of the surface
$ g $~%
\cite{GTV17}.
Thus, for the elementary square of the surface we obtain
$ d \Sigma = \sqrt{ g } d \theta d \varphi $,
where
\begin{equation}
\label{eq:sqrtg}
\sqrt{ g }
=
\frac{
- f^2 ( 1 - \sqrt{ \varepsilon } )^2 \sin \theta
\sqrt{ 1 + \sqrt{ \varepsilon } \sin ( \theta + \alpha ) + \varepsilon }
}
{ \left[ 1 + \sqrt{ \varepsilon } \sin( \theta + \alpha ) \right]^3 }.
\end{equation}
The components of the unit normal
$ \vec{ n } $
are:
\begin{equation}
\label{eq:normal}
\begin{aligned}
n_{ \rho }
&=
\frac{ \sin \theta + \sqrt{ \varepsilon } \cos \alpha }
{ \sqrt{ 1 + \sqrt{ \varepsilon } \sin ( \theta + \alpha ) + \varepsilon } }, \\
n_{ z }
&=
\frac{ \cos \theta + \sqrt{ \varepsilon } \sin \alpha }
{ \sqrt{ 1 + \sqrt{ \varepsilon } \sin ( \theta + \alpha ) + \varepsilon } }.
\end{aligned}
\end{equation}

In order to find the fields
$ \vec{ E }_{ \omega }^a $
and
$ \vec{ H }_{ \omega }^a $,
we use our combined approximate method mentioned in the Introduction and approved in recent papers~%
\cite{GTV18, TGV19}%
.
First, corresponding ``etalon'' problem should be solved to find the field in the bulk of target.
For the geometry under consideration the ``etalon'' problem is the determination of the EM field in semi-infinite medium (%
$ r_0 \le \rho < \infty $%
) with permittivity
$ \varepsilon $
having circular channel inside which the point charge
$ q $
is moving along the straight shifted trajectory.
The geometry of this problem is shown in Fig.~%
\ref{fig:geom}~%
(b).
The exact solution of this problem is presented in Appendix~%
\ref{app:etalon}.
Note that the general solution of this problem is known~%
\cite{BBol57, B57}.
However, it is more convenient to obtain this solution in the form used in~%
\cite{Ng90, Altmark05}%
, what is done in the Appendix~%
\ref{app:etalon}.

This solution in the bulk of the dielectric, Eqs.~%
\eqref{eq:Ez2}
and
\eqref{eq:Hz2}%
(we put
$ \mu = 1 $
and
$ \varphi_0 = 0 $%
), allows concluding that for
$ \rho | s | \gg 1 $
the phase term of all summands with different
$ \nu $ can be written as
\begin{equation}
\label{eq:phase}
\exp \left[ i k_0 \beta^{-1}
\left(
z + \sqrt{ \varepsilon \beta^2 - 1 } \rho
\right)  \right],
\end{equation}
i.e. it is the same as for symmetric case
$ r_0 = 0 $,
see Eq.~(5) in~%
\cite{GT17}.
This phase can be used to determine corresponding CR rays having the following angle of the refraction
$ \theta_t $:
\begin{equation}
\label{eq:sintheta}
\begin{aligned}
\sin \theta_t
&=
\frac{ - \beta^{ -1 } } { \sqrt{ r( \theta )^2 + r^{ \prime 2 }( \theta ) } }
\left[
\vphantom{ \sqrt{ \varepsilon \beta^2 - 1 } }
r( \theta ) \sin \theta - r^{ \prime }( \theta ) \cos \theta -
\right. \\
&-
\left.
\sqrt{ \varepsilon \beta^2 - 1 }
\left(
r( \theta ) \cos \theta + r^{ \prime }( \theta ) \sin \theta
\right)
\right],
\end{aligned}
\end{equation}
where
$ r^{ \prime }( \theta ) = d r( \theta ) / d \theta $%
.
Note that corresponding Eq.~(7) in~%
\cite{GT17}
contains sadly misprints.

Field
\eqref{eq:Ez2}
and
\eqref{eq:Hz2}
can be separated into two polarizations with respect to the plane of incidence determined by the incident ray and
$ \vec{ n } $%
, see Fig.~%
\ref{fig:geom}.
``Parallel'' polarization (%
$ \parallel $%
) contains components
$ E_{ z \omega } $,
$ E_{ \rho \omega } $
and
$ H_{ \varphi \omega } $.
Corresponding Fresnel coefficient is
\begin{equation}
\label{eq:Tpar}
T_{ \parallel }
=
\frac{ 2 \cos \theta_i }{ \cos \theta_i + \sqrt{ \varepsilon } \cos \theta_t }.
\end{equation}
``Ortogonal'' polarization (%
$ \bot $%
) contains components
$ H_{ z \omega } $,
$ H_{ \rho \omega } $
and
$ E_{ \varphi \omega } $.
Corresponding Fresnel coefficient is
\begin{equation}
\label{eq:Tort}
T_{ \bot }
=
\frac{ 2 \sqrt{ \varepsilon } \cos \theta_i }{ \sqrt{ \varepsilon } \cos \theta_i + \cos \theta_t }.
\end{equation}
The angle of incidence
$ \theta_i $
can be obtained out of Snell's law
$ \sqrt{ \varepsilon } \sin \theta_i = \sin \theta_t $
and
\eqref{eq:sintheta}.
The unit vector of the transmitted wave
$ \vec{ e }_k $
can be calculated as follows:
\begin{equation}
\label{eq:ek}
\begin{aligned}
e_{ k \rho } &= n_{ \rho } \cos \theta_t - n_z \sin \theta_t, \\
e_{ k z } &= n_{ \rho } \sin \theta_t + n_z \cos \theta_t.
\end{aligned}
\end{equation}

Transmitted field at the outer surface of the aperture is determined via the component of the field being orthogonal to the plane of incidence, i.e.
$ H_{ \varphi \omega } $
for
$ \parallel $%
-polarization and
$ E_{ \varphi \omega } $
for
$ \bot $%
-polarization.
In case
$ \rho_0 | s | \gg 1 $
(this means that the aperture is supposed to be far enough from the charge trajectory) at the inner side of the aperture we have
\begin{equation}
\label{eq:Hphiappr}
\begin{aligned}
&H_{ \varphi \omega }^{ a- }
=
\frac{ q \omega \exp \left( i k_0 z_0 / \beta + i \rho_0 s - 3 \pi i /  4 \right) }{ i \pi \upsilon^2 \gamma^2 }
\sqrt{ \frac{ 2 }{ \pi \rho_0 s } }
{\times} \\
&{\times}
\frac{ i k_0 }{ s^2 }
\!\!
\left\{
- \varepsilon s
I_0 ( r_0 \sigma_0 )
\tilde{ A }_0^{ ( E2 ) }
{+}
2
\sum\limits_{ \nu = 1 }^{ \infty }
I_{ \nu } ( r_0 \sigma_0 )
e^{ \frac{ i \pi ( 1 {-} \nu ) }{ 2 } }
{\times}
\right. \\
&\left.
\vphantom{ \sum\limits_{ \nu = 1 }^{ \infty } }
\times
\cos ( \nu \varphi )
\left[
\varepsilon
\tilde{ A }_{ \nu }^{ ( E2 ) }
\left( i s - \frac{ 1 }{ 2 \rho_0 } \right)
-
\frac{ i \nu }{ \beta \rho_0 }
\tilde{ A }_{ \nu }^{ ( H2 ) }
\right]
\right\},
\end{aligned}
\end{equation}
\begin{equation}
\label{eq:Ephiappr}
\begin{aligned}
&E_{ \varphi \omega }^{ a- }
=
\frac{ q \omega \exp \left( i k_0 z_0 / \beta + i \rho_0 s - 3 \pi i /  4 \right) }{ i \pi \upsilon^2 \gamma^2 }
\sqrt{ \frac{ 2 }{ \pi \rho_0 s } }
{\times} \\
&{\times}
\frac{ i k_0 }{ s^2 }
2
\sum\limits_{ \nu = 1 }^{ \infty }
I_{ \nu } ( r_0 \sigma_0 )
e^{ \frac{ i \pi ( 1 {-} \nu ) }{ 2 } }
{\times} \\
&{\times}
\sin ( \nu \varphi )
\left[
i
\tilde{ A }_{ \nu }^{ ( H2 ) }
\left( i s - \frac{ 1 }{ 2 \rho_0 } \right)
-
\frac{ \nu }{ \beta \rho_0 }
\tilde{ A }_{ \nu }^{ ( E2 ) }
\right],
\end{aligned}
\end{equation}
where notations are given in the Appendix~%
\ref{app:etalon}.
The transmitted fields at the outer surface of the aperture are
\begin{equation}
\label{eq:EphiHphi}
\begin{aligned}
H_{ \varphi \omega }^{ a }
&=
T_{ \parallel }
H_{ \varphi \omega }^{ a- },
\quad
\vec{ E }_{ \omega }^{ \parallel }
= H_{ \varphi \omega }^{ a }
\left[
\vec{ e }_{ \varphi }, \, \vec{ e }_k
\right], \\
E_{ \varphi \omega }^{ a }
&=
T_{ \bot }
E_{ \varphi \omega }^{ a- },
\quad
\vec{ H }_{ \omega }^{ \bot }
= E_{ \varphi \omega }^{ a }
\left[
\vec{ e }_k, \, \vec{ e }_{ \varphi }
\right],
\end{aligned}
\end{equation}
therefore
\begin{equation}
\label{eq:ErhozHrhoz}
\begin{aligned}
E_{ \rho \omega }^{ a }
&=
H_{ \varphi \omega }^{ a }
e_{ k z },
\quad
E_{ z \omega }^{ a }
=
- H_{ \varphi \omega }^{ a }
e_{ k \rho },
\\
H_{ \rho \omega }^{ a }
&=
- E_{ \varphi \omega }^{ a }
e_{ k z },
\quad
H_{ z \omega }^{ a }
=
E_{ \varphi \omega }^{ a }
e_{ k \rho }.
\end{aligned}
\end{equation}
%
\begin{table}[b]
\centering
\caption{Parameters of numerical calculations.}
\label{tab:pars}
\begin{tabular}{ c | c }
\hline
{ \bf Parameter }                                                        & { \bf Value }                       \\
\hline
$ \omega $                                                               & $ 2 \pi \cdot 100 $GHz              \\ \hline
$ \lambda $                                                              & $ 0.3  $cm                          \\ \hline
$ c / \omega = \lambda / ( 2 \pi ) $                                     & $ 0.05 $cm                          \\ \hline
$ \beta $                                                                & $ 0.8 $                             \\ \hline
``Focal'' parameter $ f $                                                & $ 500 c / \omega \approx 24$cm      \\ \hline
$ r( \pi ) $                                                             & $ 540 c / \omega \approx 26$cm      \\ \hline
$ q $                                                                    & $ 1 $nC                             \\ \hline
$ x_{ \max } $                                                           & $ 500 c / \omega \approx 24$cm      \\ \hline
$ x_{ \max }^{ \mathrm{ eff } } $                                        & $ 340 c / \omega \approx 16$cm      \\ \hline
$ a $                                                                    & $ c / \omega $                      \\ \hline
$ \theta_{ \min } $                                                      & $ 162^{ \circ } $                   \\ \hline
$ \theta_{ \max } $                                                      & $ 179^{ \circ } $                   \\ \hline
$ \varepsilon $                                                          & $ 1.6 $                             \\ \hline
$ \alpha $                                                               & $ 80^{ \circ } $                    \\ \hline
$ \alpha_{ \mathrm{ tar } } $                                            & $ 73^{ \circ } $                    \\ \hline
\end{tabular}
\end{table}

Eqs.~%
\eqref{eq:EphiHphi}
and
\eqref{eq:ErhozHrhoz}
show that the field distribution over the outer surface of the aperture is defined and all the things needed for the evaluation of the integral~%
\eqref{eq:SChu}
are ready.

%
\begin{figure*}
\centering
\includegraphics[width=0.95\linewidth]{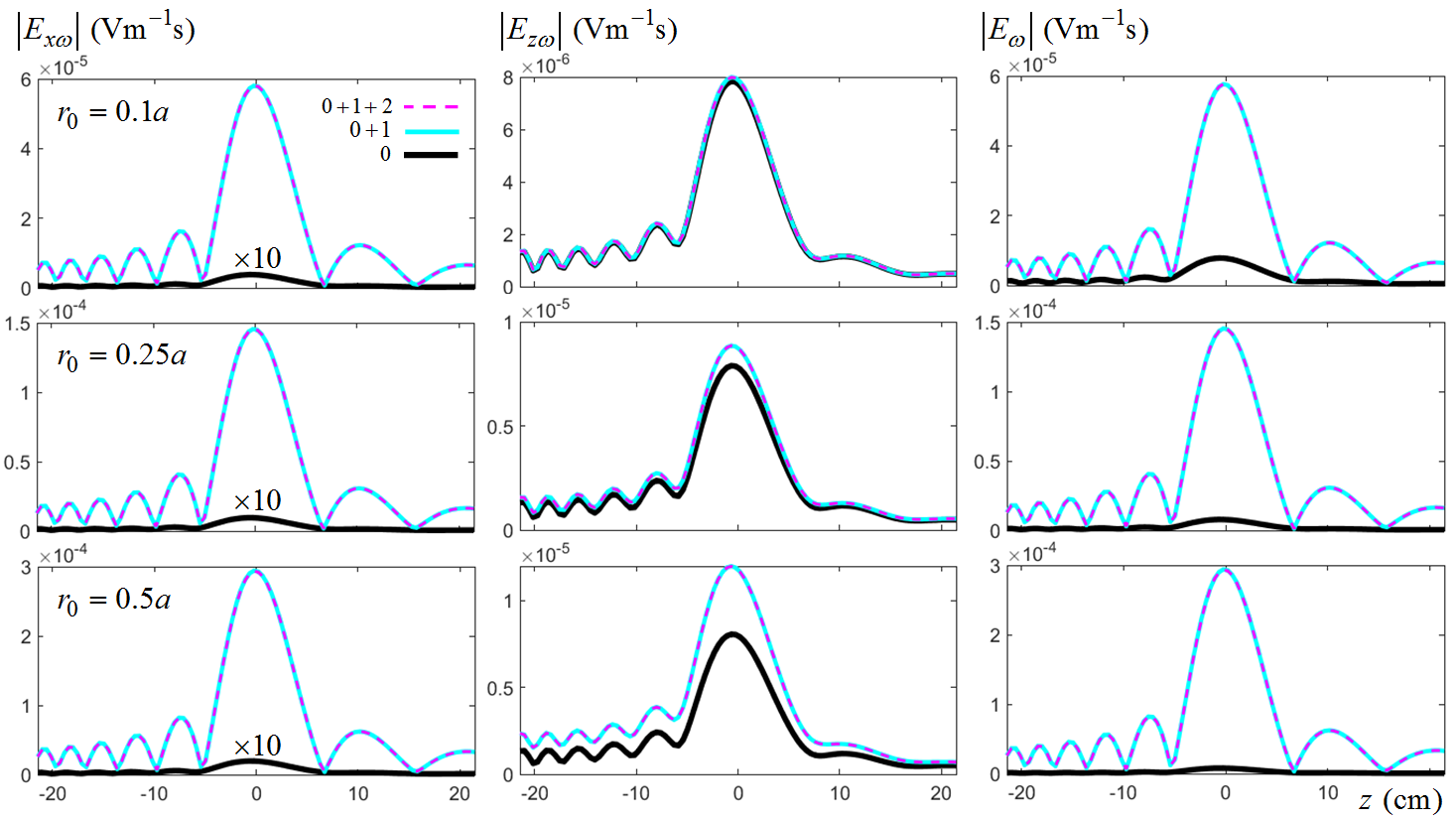}
\caption{\label{fig:Ealongz}%
Field behavior along the line parallel to
$ z $%
-axis and having offset
$ r_0 $
in
$ x $%
-direction.
Black (thickest) line (multiplied by factor
$ 10 $
in the left column to be recognized from zero) corresponds to the contribution of the term with
$ \nu = 0 $
in Eq.~%
\eqref{eq:Hphiappr}.
Cyan (thinner) and magenta (dashed) lines correspond to the contribution of two terms (with
$ \nu = 0, 1 $%
) and three terms (with
$ \nu = 0, 1, 2 $%
), correspondingly, in Eqs.~%
\eqref{eq:Hphiappr}
and
\eqref{eq:Ephiappr}.
Top row is for
$ r_0 = a / 10 $,
middle row is for
$ r_0 = a / 4 $
and bottom row is for
$ r_0 = a / 2 $.
Note that black line on the second plot in the first line is present but partially covered by the other ones.
}
\end{figure*}
%

Prior presenting numerical results, let us discuss the obtained analytical results.
The ``mode'' with number
$ \nu = 0 $
is manifested in
$ \parallel $%
-polarization only.
The corresponding
$ \varphi $%
-independent term in Eq.~%
\eqref{eq:Hphiappr}
is equal to the one in the case with non-shifted charge multiplied by a factor
$ I_0 ( r_0 \sigma_0 ) < 1 $.
Therefore, due to this zeroth ``mode'' the same effect of concentration will take place in non-symmetrical case (with corresponding scaling of the field).
In particular, the transverse electric field produced by the concentrator in this term is exact zero for
$ \rho = 0 $~%
\cite{GTV17}.
However, ``modes'' with numbers
$ \nu \ge 1 $
will add
$ \varphi $%
-dependent field distribution over the aperture, therefore, altering this result.
In Sec.~%
\ref{sec:num}
we will clarify (using corresponding numerical examples) the questions concerning the required number of asymmetric ``modes'' to be taken into account and their influence on the field distribution near the focal plane.

\section{\label{sec:num}Numerical results}

Here we present results of EM field calculation in the area outside the target (mainly, near the focal point which is of the most interest) using Eq.~%
\eqref{eq:SChu}.
The limits of integration over
$ \varphi $
are $ \left( 0, 2 \pi \right) $,
while the limits of integration over
$ \theta $,
$ \theta_{ \min } $
and
$ \theta_{ \max } $%
, are determined by dimensions of the concentrator.

%
\begin{figure*}
\centering
\includegraphics[width=0.9\linewidth]{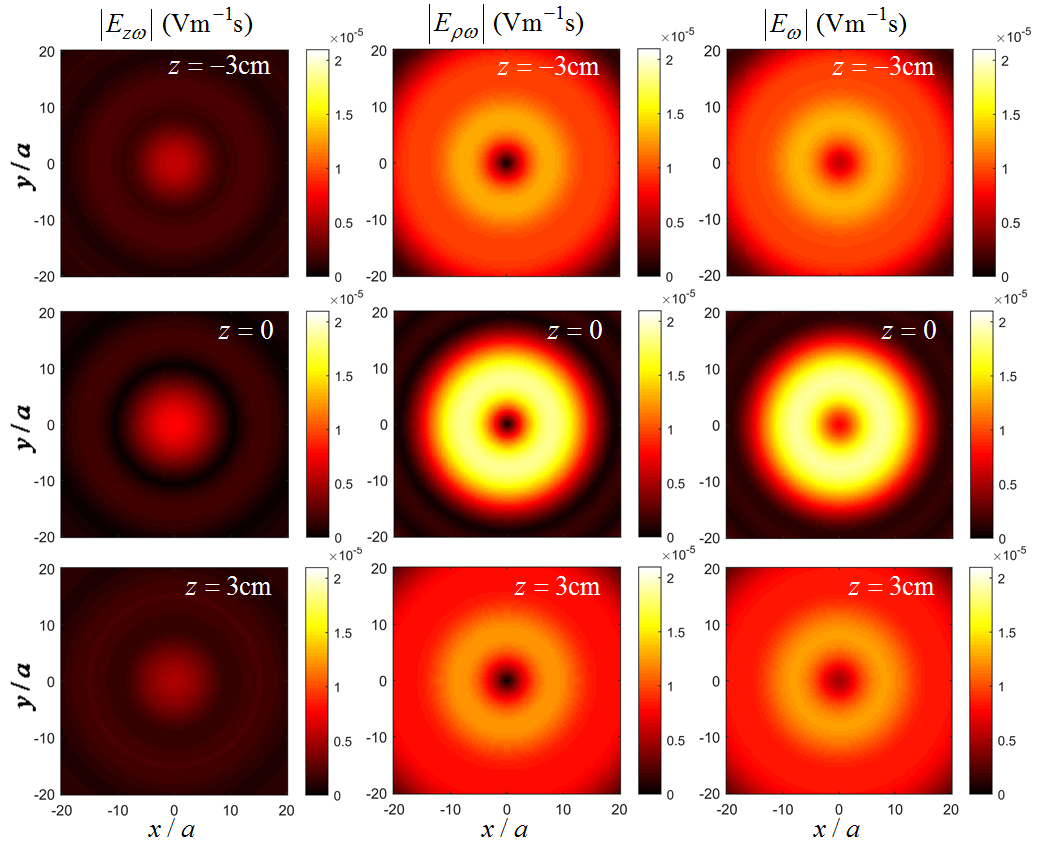}
\caption{\label{fig:FieldsSymm}%
Two-dimensional distribution of the absolute values of the longitudinal (%
$ E_{ z \omega } $%
), the transverse (%
$ E_{ \rho \omega } $%
) and the total (%
$ E_{ \omega } $%
) field over
$ xy $%
-plane for symmetrical case
$ r_0 = 0 $.
Plane
$ z = z_f = 0 $
is the focal plane of the concentrator.
Calculation parameters are given in Table~%
\ref{tab:pars}.
}
\end{figure*}
%

%
A numerical code was realized in MATLAB with the use of Parallel Computing Toolbox for evaluation of integrals~%
\eqref{eq:SChu}.
An approximate time consumed for the calculation of each subplot (%
$ 100 \times 100 $
points) in Fig.~%
\ref{fig:FieldsSymm}
was in the range
$ 50 - 250 $
seconds per CPU core depending on the model of processor.

The set of parameters that were used for numerical calculations is presented in Table~%
\ref{tab:pars}.
First, we have put
$ z_f = 0 $,
i.e. position of the designed focus is
$ z = 0 $.
Then,
we have determined all the dimensions in units of
$ c / \omega $ for the chosen frequency
$ \omega $.
The channel radius was chosen to be
$ a = c / \omega $
and this radius determines
$ \theta_{ \max } $.
Dielectric permittivity
$ \varepsilon $
was chosen so that
$ \alpha $
was relatively large.
In this case the rays inside the target propagate with relatively small angle with respect to
$ z $%
-axis and target longitudinal and transverse dimensions are comparable (contrary to the case of relatively small
$ \alpha $
where the target is strongly prolonged~%
\cite{GTVGBproc16}%
).

Transverse dimensions of the target are determined by
$ x_{ \max } $,
see Fig~%
\ref{fig:geom}~%
(a).
Back slope of the target is cut by straight line determined by
$ \alpha_{ \mathrm{ tar } } $
which should be smaller than
$ \alpha $.
In this case, the ray originating from the most distant (from the focus) point of the target will reach the aperture.
This ray is shown in Fig.~%
\ref{fig:geom}~%
(a).
This ray also determines the ``effective'' transverse size of the target
$ x_{ \max }^{ \mathrm{ eff } } $
which is smaller than
$ x_{ \max } $.
This point is taken into account in calculations:
upper limit
$ \theta_{ \max } $
is determined by
$ x_{ \max }^{ \mathrm{ eff } } $.

Figure~%
\ref{fig:Ealongz}
illustrates dependence of absolute values of
$ E_{ x \omega } $,
$ E_{ z \omega } $
and
$ E_{ \omega } $
on
$ z $
for
$ x = r_0 $
and
$ y = 0 $
(along straight line parallel to
$ z $%
-axis) and for three values of
$ r_0 = a / 10, \, a / 4, \, a / 2 $.
Black (thickest) line shows field generated by
$ 0 $%
th (symmetrical with respect to
$ \varphi $%
) ``mode'' in Eq.~%
\eqref{eq:Hphiappr}.
Recall that this ``mode'' gives the same field as in symmetrical case (only scaled by the factor
$ I_0 ( r_0 \sigma_0 ) $%
).
One can see that transverse field
$ E_{ x \omega } $
is negligible near
$ z $%
-axis, which is natural due to the symmetry considerations, while longitudinal field
$ E_{ z \omega } $
is dominant.
These are results obtained earlier~%
\cite{GTVGBproc16,GTV17}.

Cyan (thinner) line in Fig.~%
\ref{fig:Ealongz}
shows field generated by two those ``modes'' (%
$ 0 $%
th and
$ 1 $%
st), while dashed magenta line shows field generated by three those ``modes'',
$ 0 $%
th,
$ 1 $%
st and
$ 2 $%
nd.
As one can see, even for
$ r_0 = a / 2 $
contribution of the 2nd ``mode'' is negligible, therefore we can restrict ourselves by taking into account mentioned three modes in Eqs.~%
\eqref{eq:Hphiappr}
and
\eqref{eq:Ephiappr}.

It should be noted that EM field of the
$ \nu $%
-th ``mode'' is proportional to the term
$ I_{ \nu } ( r_0 \sigma_0 ) $.
In our case (see Table~%
\ref{tab:pars}%
), the argument of the modified Bessel function for the largest considered offset
$ r_0 = a / 2 = \lambda / ( 4 \pi ) $
is
$ r_0 \sigma_0 = \sqrt{ 1 - \beta^2 } / ( 2 \beta ) = 0.375 $%
, therefore it can be considered as small for
$ 0 < r_0 < a / 2 $.
This leads to the conclusion that each ``mode'' is proportional (in the leading order) to
$ ( r_0 \sigma_0 )^{ \nu } $
and, therefore, contributions of high-order ``modes'' decrease with
$ \nu $%
, this fact has been illustrated above in Fig.~%
\ref{fig:Ealongz}%
.
It should be underlined that the above conclusion on the number of ``modes'' required for correct EM field description
is connected mainly with the small channel radius
$ a $
in comparison with the wavelength
$ \lambda $.
If we consider channel with radius compared or larger than the wavelength this result will be essentially altered.

Moreover, if
$ a $
is much larger than the wavelength, i.e.
$ \sigma_0 a \gg 1 $%
, then
$ | A_{ \nu }^{ ( E,H2 ) } | \sim \exp( - \sigma_0 a ) $
and EM field in the bulk of the target is exponentially small for small offsets, i.e. for
$ \sigma_0 r_0 \lesssim 1 $.
However, for the offset comparable with channel radius (%
$ \sigma_0 r_0 \sim \sigma_0 a \gg 1 $%
) the field can be effectively generated.
In this case, the question on the number of required ``modes'' should be considered separately.

In order to illustrate the main results we will plot two-dimensional field distributions.
Figure~%
\ref{fig:FieldsSymm}
shows two-dimensional distribution of the longitudinal (%
$ E_{ z \omega } $%
) and the transverse (%
$ E_{ \rho \omega } $%
) components over
$ xy $%
-plane for the symmetric case
$ r_0 = 0 $.
Note that this is not a principally new result but a new illustration of the previous results obtained in~%
\cite{GTV17, GTV18}.
As one can clearly see from Fig.~%
\ref{fig:FieldsSymm}%
, transverse field is exact zero on the symmetry axis while the longitudinal field is maximal here.
The transverse field quickly increases while observation point is shifted from
$ z $%
-axis and has its maximum value around two times larger compared to maximum of
$ E_{ z \omega } $.
In all cases, fields strongly decrease with an increase in shift of the observation plane from the focal plane
$ z = 0 $.
The total field is practically determined by transverse field excluding the mentioned central area (a circle with radius
$ \approx 3 a $%
) where longitudinal field is dominant.

\begin{figure*}
\centering
\includegraphics[width=0.9\linewidth]{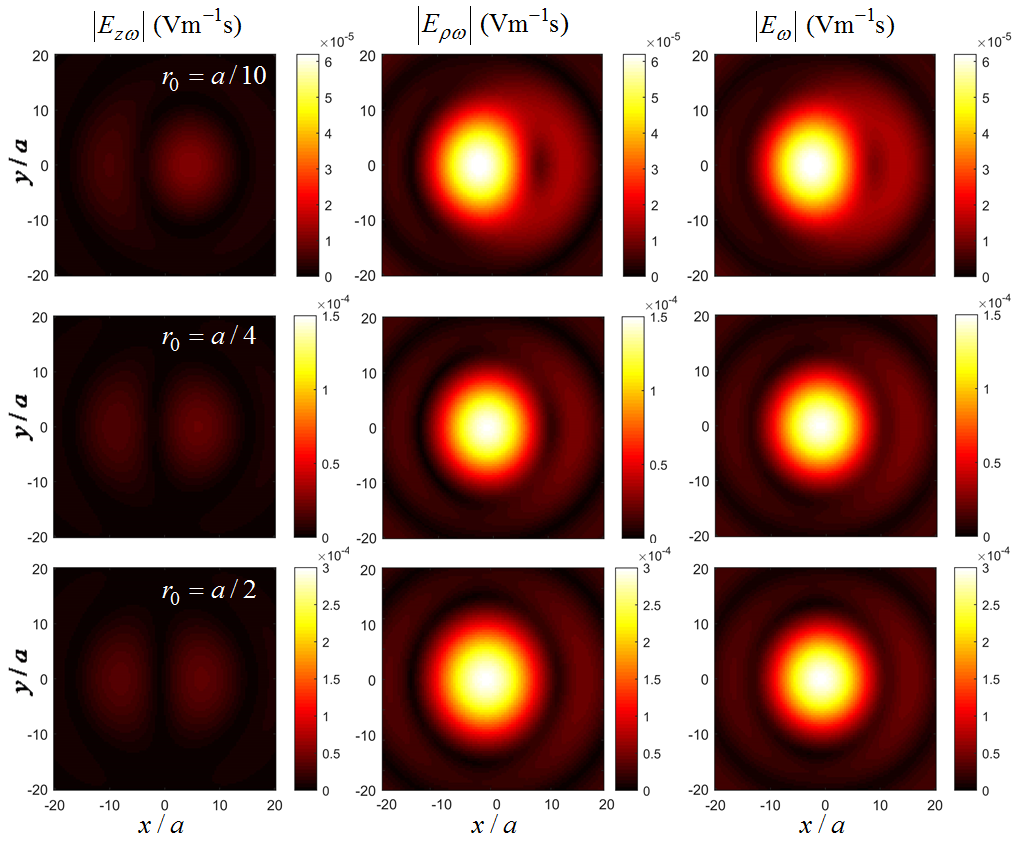}
\caption{\label{fig:FieldsAsymm}%
Two dimensional distribution of absolute values of longitudinal (%
$ E_{ z \omega } $%
) and transverse (%
$ E_{ \rho \omega } $%
) and total field over
$ xy $%
-plane in the focal plane of the concentrator (%
$ z = 0 $%
) for three values of charge's shift
$ r_0 = a / 10, \, a / 4, \, a / 2 $.
Calculation parameters are given in Table~%
\ref{tab:pars}.}
\end{figure*}

The field distribution for the case with a charge shifted from the
$ z $%
-axis is shown in Fig.~%
\ref{fig:FieldsAsymm}
for the focal plane
$ z = z_f = 0 $.
Note that tendencies in field behaviour connected with the shift of the observation plane from the focal plane (%
$ z < 0 $
or
$ z > 0 $%
) are practically the same as for symmetrical case.
The magnitudes of transverse and longitudinal fields also differ approximately by the factor
$ 5 $
and increase almost linearly with increasing
$ r_0 $.

\begin{figure}
\centering
\includegraphics[width=0.85\linewidth]{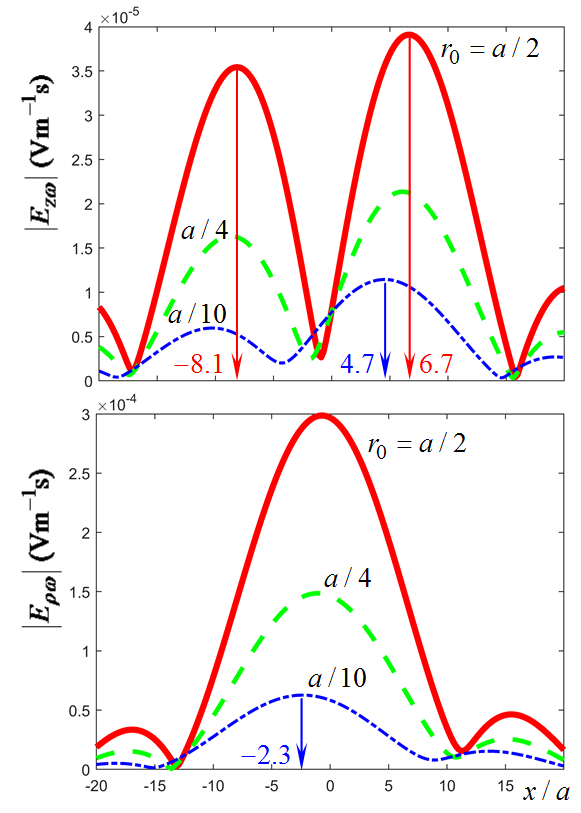}
\caption{\label{fig:peaks}%
Dependence of absolute values of longitudinal (%
$ E_{ z \omega } $%
) and transverse (%
$ E_{ \rho \omega } $%
) field over
$ x $
for
$ y = z = 0 $
for three values of charge's shift
$ r_0 = a / 10, \, a / 4, \, a / 2 $.
Calculation parameters are the same as in Fig.~%
\ref{fig:FieldsAsymm}%
.}
\end{figure}

The most interesting feature is occurred in longitudinal field (see Fig.~%
\ref{fig:FieldsAsymm}%
, left column).
For the relatively small offset (%
$ r_0 = a / 10 $%
) this field is strongly asymmetric with respect to
$ x = 0 $.
The stronger peak is located in the area
$ x > 0 $%
, i.e. in the area where the shifted charge propagates (recall that
$ \varphi_0 = 0 $%
, therefore the charge is shifted to the positive
$ x $%
).
With an increase in
$ r_0 $
these peaks become more symmetrical (%
$ r_0 = a / 4 $%
), and for relatively large offset (%
$ r_0 = a / 2 $%
) we have two almost symmetrical peaks located approximately at
$ x =  6.7 a $
and
$ x =  -8.7 a $%
, i.e. far enough from the charge trajectory.
More detailed position of the discussed peaks is shown in Fig.~%
\ref{fig:peaks} .

In turn, the transverse field (see Fig.~%
\ref{fig:FieldsAsymm}%
, second column) also has an essential asymmetry for relatively small offsets (%
$ r_0 = a / 10 $%
).
It is worth noting that the peak of transverse field is shifted in opposite direction compared to the shift of the charge trajectory, its approximate location is
$ x = -2.3 a $
(see Fig.~%
\ref{fig:peaks}
for details).
Therefore in this case (small offset) one can potentially detect these peaks separately: the peak of
$ E_{ z \omega } $
for
$ x > 0 $
and the peak of
$ E_{ \rho \omega } $
for
$ x < 0 $.
For larger offsets, the peak of
$ E_{ \rho \omega } $
becomes practically symmetric in both
$ x $
and
$ y $
direction, is located nearly in the center and totally dominates the longitudinal peak.

%
%

%
%

\section{\label{sec:discuss}Discussions}

Possible applications of the presented dielectric concentrator would lie in the area of beam diagnostics and beam manipulation.
Since strong field concentration takes place in the focal plane near the focal point, sensitivity and accuracy of mentioned diagnostics can be essentially increased.
For example, peculiarities of field distribution for
$ r_0 \ne 0 $
can be used for determination of beam shift and positioning of the beam toward the axis of the structure.
Note that the difference between magnitudes of two peaks of top plot in Fig.~%
\ref{fig:peaks}
depends on
$ r_0 $.
For
$ r_0 = a / 10 $
left peak magnitude is about 50\% of the right peak magnitude while for
$ r_0 = a / 2 $
left peak magnitude is about 90\% of the right peak magnitude.
Moreover, these peaks are shifted for several values of
$ a $
(in our case from
$ \approx 5 $
to
$ \approx 8 $%
) from the axis, thus, simplifying the detection of the field.
However, to realize this diagnostic scheme one should separate the longitudinal field from the much more intensive transverse field what is rather challenging.
On the other hand, strong peak of the transverse field can be utilized for the same purpose if we are interested in relatively small offsets.
As it is seen from the bottom plot in Fig.~%
\ref{fig:peaks}%
, for
$ r_0 = a / 10 $
the peak is located at
$ x = -2.3 a $
which is far enough from the structure axis and suitable for detection.%

If the bunch is well aligned along the axis (%
$ r_0 = 0 $%
), it experiences influence of the longitudinal field (%
$ E_{ z \omega } $%
) only, see Fig.~%
\ref{fig:FieldsSymm}.
Interaction between the bunch and strongly concentrated radiated field can lead to longitudinal modulation of the bunch.
As it has been experimentally shown in~%
\cite{Ant13}%
, a 2-inch-long dielectric-lined capillary can produce energy modulation with approximately
$ 0.25 $~%
MeV magnitude which corresponds to
$ 5~\mathrm{MV/m} $
longitudinal field magnitude inside the capillary.
In the case shown in Fig.~%
\ref{fig:FieldsSymm}, an interaction area (focal spot) can be approximately estimated to be
$ 6 $~%
cm-long.
Moreover, as was shown in Ref.~%
\cite{GTV18}%
, if typical energy spread inside the bunch (about
$ 0.5 $\%%
) is taken into account, the focal spot length is increased up to 2.5 times.
Therefore, interaction area in the case of concentrator is comparable or larger than that in Ref.~%
\cite{Ant13}%
.
As our additional calculations show,
a typical value of Fourier harmonic magnitude of the longitudinal field
$\approx 10^{ -5 }~\mathrm{MV/m \cdot s} $
is achieved, at least, in the frequency range
$ 100 - 200 $~%
GHz. Therefore, we can approximately estimate the peak magnitude of the radiation pulse (produced by this frequency range) in the focal area by a simple multiplication of the mentioned Fourier harmonic magnitude by the factor
$ 2 \pi \cdot 100~\mathrm{GHz} $.
This results in approximately
$ 6 \mathrm{MV/m} $
longitudinal field magnitude and leads to a conclusion that the discussed bunch modulation is potentially achievable.

\section{\label{sec:concl}Conclusion}

In the present paper, we have presented analytical and numerical investigation of the EM radiation produced by a point charge moving through the dielectric concentrator for Cherenkov radiation (see~%
\cite{GT14}%
) for the case of off-axis charge movement.
The problem has been solved using original approach (developed and verified in our previous papers) suitable for investigation of radiation from various dielectric targets of finite size with several sharp interfaces.
It is worth noting that this approach possesses asymptotic accuracy with respect to the ratio between wavelength and transverse size of the target.

Contrary to the symmetric case considered in our papers earlier, here the strong asymmetry of the field distribution near the focal plane is observed and illustrated.
For example, with an increase in charge offset, peak of longitudinal field is first shifted in the direction of charge shift and then is divided into two peaks.
Since the difference between magnitudes of these peaks depend on charge offset, detection of these peaks can be possibly used for determination of the charge offset and the charge alignment inside the structure.
Note that transverse position of the discussed peaks is several channel radii, therefore they can be potentially detected without a need to deflect or damp the charge.

\section{Acknowledgments}

This work was supported by Russian Foundation for Basic Research (RFBR), grant No.~17-52-04107.

\appendix

\section{\label{app:concdetails} Detailed explanation of the outer surface of the concentrator for CR }

%
\begin{figure}
\centering
\includegraphics[width=0.95\linewidth]{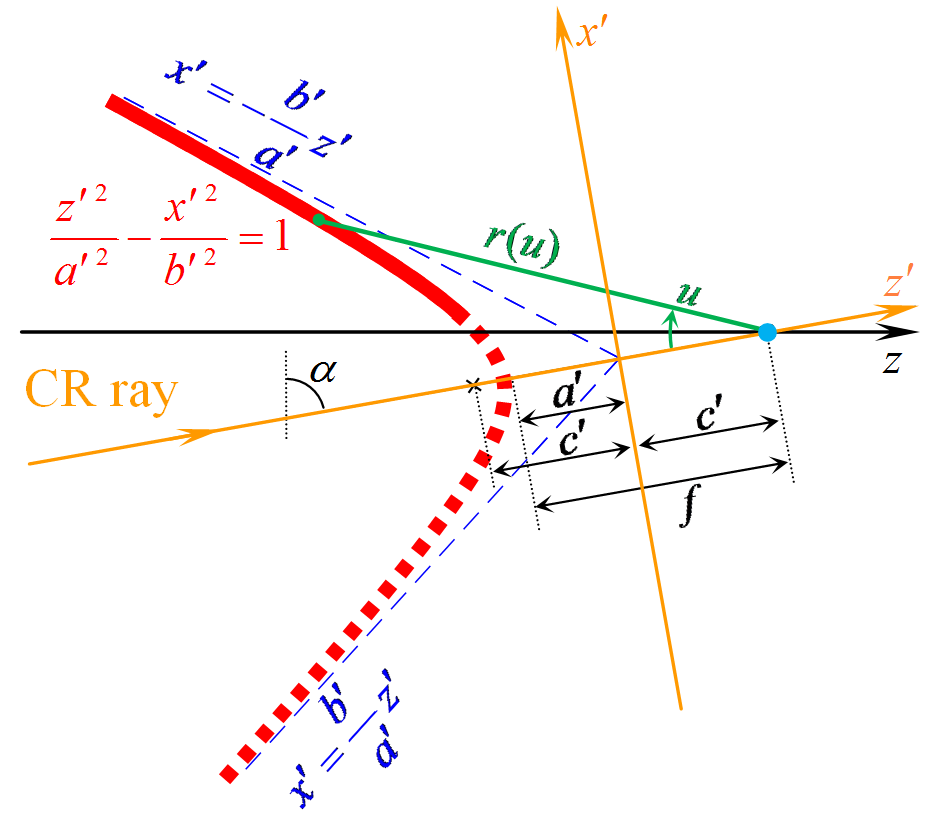}
\caption{\label{fig:geom_details}%
Detailed description of the outer surface of the concentrator for CR.}
\end{figure}
%

Figure~%
\ref{fig:geom_details}
explains the determination of the surface shape of the target under consideration in more detail.
Let us imagine a single ray of CR directed to the focus point.
This ray is inclined with respect to
$ z $%
-axis by the angle
$ \pi / 2 - \alpha $%
, where
$ \alpha = {\rm arcsin }[ 1 / ( \sqrt{ \varepsilon } \beta ) ] $%
.
In the Cartesian coordinate frame
$ ( z^{ \prime}, x^{ \prime } ) $
associated with this ray, the outer surface profile corresponds to a branch of hyperbola (red line):
\begin{equation}
\label{eq:hyper}
\left(
\frac{ z^{ \prime } }{ a^{ \prime } }
\right)^2
-
\left(
\frac{ x^{ \prime } }{ b^{ \prime } }
\right)^2
=
1,
\end{equation}
where the standard hyperbola parameters are:
\begin{equation}
\label{eq:hyperpars}
a^{ \prime } = \frac{ f }{ \sqrt{ \varepsilon } + 1 },
\quad
b^{ \prime } = f \sqrt{ \frac{ \sqrt{ \varepsilon } - 1 }{ \sqrt{ \varepsilon } - 1 } },
\quad
c^{ \prime } = \frac{ f \sqrt{ \varepsilon } }{ \sqrt{ \varepsilon } + 1 },
\end{equation}
while
$ ( c^{ \prime } )^2 = ( a^{ \prime } )^2 + ( b^{ \prime } )^2 $
and
$ f = a^{ \prime } + c^{ \prime } $.
Recall that
$ a^{ \prime } $
is the semi major axis,
$ c^{ \prime } $
is the distance from the coordinate frame origin to the focus and the ratio
$ b^{ \prime } / a^{ \prime } $
determines the asymptotes.
The focus lying oh the
$ z $%
-axis is the focus of concentrator.
If we put the origin of the polar coordinate frame
$ ( r, u ) $
in this focus we obtain the following standard parametrization for the discussed branch of hyperbola:
\begin{equation}
\label{eq:hyperpolar}
r( u )
=
\frac{ p^{ \prime } }{ \epsilon^{ \prime } \cos u - 1 }
=
\frac{ f \left( \sqrt{ \varepsilon } - 1 \right) }{ \sqrt{ \varepsilon } \cos u - 1 },
\end{equation}
where
$ p^{ \prime } = ( b^{ \prime } )^2 / a^{ \prime } $
and
$ \epsilon^{ \prime } = c^{ \prime } / a^{ \prime } > 1 $
is the eccentricity.
If we take into account simple relation between
$ u $
and
$ \theta $,
\[
u = \frac{ 3 \pi }{ 2 } - ( \alpha + \theta ),
\]
we arrive at Eq.~%
\eqref{eq:rTheta}.
Finally, in order to obtain the full outer surface of the concentrator shown in Fig.~%
\ref{fig:geom}%
, we should cut the piece of Eq.~%
\eqref{eq:rTheta}
determined by the range
$ \left[ \theta_{ \min }, \theta_{ \max } \right] $
(this peace is shown by solid red in Fig.~%
\ref{fig:geom_details}%
) and rotate this piece over
the
$ z $%
-axis.

\section{\label{app:etalon} Solution of ``etalon'' problem }

For the case under consideration, geometry of ``etalon'' problem is shown in Fig.~%
\ref{fig:geom}~%
(b).
Within this section, cylindrical frame (%
$ \rho $,
$ \varphi $
$ z $%
) is utilized.
One should find the EM field in semi-infinite medium (%
$ a \le \rho < \infty $%
) with permittivity
$ \varepsilon $
and permeability
$ \mu $
having a cylindrical channel where a charge propagates along shifted trajectory.
It is supposed that the trajectory is shifted in
$ \varphi = \varphi_0 $
direction and
$ r_0 $
is the value of this shift.
In addition, we will refer to the area
$ \rho \le a $
as ``area 1'' and to the area
$ \rho \ge a $
as ``area 2''.
As was mentioned above, despite of the fact that general solution of this problem is known~%
\cite{BBol57, B57}
it is useful to obtain it in more convenient form, similar to that in~%
\cite{Ng90, Altmark05}.

The problem is solved for Fourier transforms of field components, for example:
\begin{equation}
\label{eq:Fourier}
\vec{ E }_{ \omega }
=
\frac{ 1 }{ 2 \pi }
\int\nolimits_{ - \infty }^{ + \infty }
\vec{ E } \exp( i \omega t ) \, dt.
\end{equation}
Charge and current densities have the form
\begin{equation}
\label{eq:rhoj}
\rho_q
=
q \frac{ \delta( \rho - r_0 ) }{ \rho }
\delta( \varphi - \varphi_0 )
\delta( z - \upsilon t ),
\,\,
j = \rho_q \upsilon,
\,\,
\vec{ j } = j \vec{ e }_z.
\end{equation}
Here
$ \upsilon = \beta c $
and
$ \delta( \varphi - \varphi_0 ) $
means
$ 2 \pi $%
-periodic function with corresponding
$ \delta $
singularity in each period.
This function can be decomposed in Fourier series as follows:
\begin{equation}
\label{eq:deltaphidecomp}
\delta( \varphi - \varphi_0 )
=
\frac{ 1 }{ 2 \pi }
\sum\nolimits_{ \nu = - \infty  }^{ + \infty }
\exp( - i \nu ( \varphi - \varphi_0 ) ).
\end{equation}
For radial part in~%
\eqref{eq:rhoj}
the following decomposition holds for arbitrary integer
$ \nu $~%
\cite{KKb}:
\begin{equation}
\label{eq:deltardecomp}
\frac{ \delta( \rho - r_0 ) }{ \rho }
=
\int\nolimits_0^{ + \infty }
J_{ \nu }( \rho \xi ) J_{ \nu }( r_0 \xi ) \xi d\xi,
\end{equation}
where
$ J_{ \nu } $
is Bessel function of order
$ \nu $.
Fourier transforms~%
\eqref{eq:Fourier}
of the sources~%
\eqref{eq:rhoj}
have the form:
\begin{equation}
\label{eq:rhojFourier}
\rho_{ q \omega }
=
\frac{ q }{ 2 \pi \upsilon }
\frac{ \delta ( \rho - r_0 ) }{ \rho }
\delta( \varphi - \varphi_0 )
\exp \left(  i \frac{ \omega z }{ \upsilon } \right),
\,\,
j_{ \omega } = \upsilon \rho_{ q \omega }.
\end{equation}

From Maxwell's equations for magnitudes of Fourier harmonics
\begin{equation}
\label{eq:Maxwell}
\begin{aligned}
\mathrm{ rot } \vec{ E }_{ \omega }
&=
i k_0 \mu \vec{ H }_{ \omega }, \\
\mathrm{ rot } \vec{ H }_{ \omega }
&=
- i k_0 \varepsilon \vec{ E }_{ \omega }
+
\frac{ 4 \pi }{ c } \vec{ j }_{ \omega }, \\
\varepsilon \mathrm{ div } \vec{ E }_{ \omega }
&=
4 \pi \rho_{ q \omega },
\quad
\mathrm{ div } \vec{ H }_{ \omega }
=
0,
\end{aligned}
\end{equation}
$ k_0 = \omega / c $%
, one can obtain the following two equations for longitudinal components of electric and magnetic field~%
\cite{Ng90, Altmark05}:
\begin{equation}
\begin{aligned}
\label{eq:Helmholtz}
\Delta E_{ z \omega } + k_0^2 \varepsilon \mu E_{ z \omega }
&=
\frac{ 4 \pi }{ \varepsilon }
\frac{ \partial \rho_{ q \omega } }{ \partial z }
-
\frac{ 4 \pi i \omega \mu }{ c^2 }
j_{ \omega }, \\
\Delta H_{ z \omega } + k_0^2 \varepsilon \mu H_{ z \omega }
&=
0,
\end{aligned}
\end{equation}
where
\begin{equation}
\label{eq:Laplace}
\Delta
=
\frac{ \partial^2 }{ \partial \rho^2 } + \frac{ 1 }{ \rho } \frac{ \partial }{ \partial \rho }
+
\frac{ 1 }{ \rho^2 } \frac{ \partial }{ \partial \varphi^2 }
+
\frac{ \partial^2 }{ \partial z^2},
\end{equation}
and the following expressions for the rest of the field components:
\begin{equation}
\label{eq:ErHrFourier}
\begin{aligned}
E_{ \rho \omega }
&=
\frac{ i }{ k_0 \varepsilon }
\left(
\frac{ 1 }{ \rho }
\frac{ \partial H_{ z \omega } }{ \partial \varphi }
-
\frac{ \partial H_{ \varphi \omega } }{ \partial z }
\right), \\
H_{ \rho \omega }
&=
\frac{ - i }{ k_0 \mu }
\left(
\frac{ 1 }{ \rho }
\frac{ \partial E_{ z \omega } }{ \partial \varphi }
-
\frac{ \partial E_{ \varphi \omega } }{ \partial z }
\right),
\end{aligned}
\end{equation}
\begin{equation}
\label{eq:HphiEphiFourier}
\begin{aligned}
\left(
\frac{ \partial^2 }{ \partial z^2 }
+
k_0^2 \varepsilon \mu
\right)
H_{ \varphi \omega }
&=
\frac{ 1 }{ \rho }
\frac{ \partial^2 H_{ z \omega } }{ \partial z \partial \varphi }
+
i k_0 \varepsilon
\frac{ \partial E_{ z \omega } }{ \partial \rho }, \\
\left(
\frac{ \partial^2 }{ \partial z^2 }
+
k_0^2 \varepsilon \mu
\right)
E_{ \varphi \omega }
&=
\frac{ 1 }{ \rho }
\frac{ \partial^2 E_{ z \omega } }{ \partial z \partial \varphi }
-
i k_0 \mu
\frac{ \partial H_{ z \omega } }{ \partial \rho }.
\end{aligned}
\end{equation}
Since
$ z $%
-dependence in sources~%
\eqref{eq:rhojFourier}
is
$ \exp( i \omega z / \upsilon ) $,
field components should have the same dependence, i.e.:
\begin{equation}
\label{eq:tildeFourier}
\left\{
\vec{ E }_{ \omega },
\,
\vec{ H }_{ \omega }
\right\}
=
\left\{
\vec{ \tilde{ E } }_{ \omega },
\,
\vec{ \tilde{ H } }_{ \omega }
\right\}
\exp \left( i \frac{ \omega z }{ \upsilon } \right),
\end{equation}
therefore
$ \partial / \partial z \to i k_0 / \beta $
in Eqs.~%
\eqref{eq:Helmholtz},
\eqref{eq:Laplace},
\eqref{eq:ErHrFourier}
and
\eqref{eq:HphiEphiFourier}.
Moreover, since
$ \vec{ \tilde{ E } }_{ \omega } $
should be
$ 2 \pi $
periodic function, it should be decomposed in Fourier series, similar to~%
\eqref{eq:deltaphidecomp}:
\begin{equation}
\label{eq:tildeFourierphi}
\vec{ \tilde{ E } }_{ \omega }
=
\sum\nolimits_{ \nu = - \infty }^{ + \infty }
\vec{ \tilde{ E } }_{ \omega \nu }
\exp ( - i \nu \varphi ),
\end{equation}
therefore
$ \partial / \partial \varphi \to - i \nu $
in Eqs.~%
\eqref{eq:Helmholtz},
\eqref{eq:Laplace},
\eqref{eq:ErHrFourier}
and
\eqref{eq:HphiEphiFourier}.
Taking into account all these considerations, we obtain from Eq.~%
\eqref{eq:Helmholtz}
\begin{equation}
\label{eq:Helmholtsdecomposed}
\begin{aligned}
\left[
\frac{ \partial^2 }{ \partial \rho^2 } + \frac{ 1 }{ \rho } \frac{ \partial }{ \partial \rho }
+
s^2
-
\frac{ \nu ^2 }{ \rho^2 }
\right]
\tilde{ E }_{ z \omega \nu }
&=
4 \pi \tilde{ Q }_{ \omega \nu }, \\
\left[
\frac{ \partial^2 }{ \partial \rho^2 } + \frac{ 1 }{ \rho } \frac{ \partial }{ \partial \rho }
+
s^2
-
\frac{ \nu ^2 }{ \rho^2 }
\right]
\tilde{ H }_{ z \omega \nu }
&= 0,
\end{aligned}
\end{equation}
\begin{equation}
\label{eq:tildeQ}
\tilde{ Q }_{ \omega \nu }
=
\frac{ i q \sigma^2 }{ 4 \pi^2 \omega \epsilon }
e^{ i \nu \varphi_0 }
\int\nolimits_0^{ + \infty }
J_{ \nu }( \rho \xi ) J_{ \nu }( r_0 \xi ) \xi d\xi,
\end{equation}
\begin{equation}
\label{eq:ssigma}
s^2 = \frac{ k_0^2 }{ \beta^2 }( \varepsilon \mu \beta^2 - 1 ),
\quad
\sigma^2 = - s^2.
\end{equation}

Particular solution (this solution will be supported by upper index
$ ( i ) $
and will be further referred to as ``incident'' field) of Eq.~%
\eqref{eq:Helmholtsdecomposed}
is
$ \tilde{ H }_{ z \omega \nu }^{ ( i ) } = 0 $,
\begin{equation}
\label{eq:Ezincidentdecomposed}
\tilde{ E }_{ z \omega \nu }^{ ( i ) }
=
\frac{ i q \sigma^2 }{ \pi \omega \epsilon }
e^{ i \nu \varphi_0 }
\int\nolimits_0^{ + \infty }
\frac{ J_{ \nu }( \rho \xi ) J_{ \nu }( r_0 \xi ) }
{ s^2 - \xi^2 } \xi d\xi.
\end{equation}

We will further suppose that ``incident'' field~%
\eqref{eq:Ezincidentdecomposed}
is determined inside the vacuum channel (area 1,
$ \rho \le a $%
), where
$ \varepsilon = \mu = 1 $,
$ \sigma^2 = \sigma_0^2 = k_0^2 \beta^{ - 2 } ( 1 - \beta^2 )  = k_0^2 \beta^{ - 2 } \gamma^{ - 2 } $,
$ \gamma $
is Lorentz factor.
Integral over
$ \xi $
in~%
\eqref{eq:Ezincidentdecomposed}
is tabular~%
\cite{PBMb2},
and we obtain:
\begin{equation}
\label{eq:Ezincident}
\tilde{ E }_{ z \omega \nu }^{ ( i ) }
=
\frac{ - i q \sigma_0^2 }{ \pi \omega }
e^{ i \nu \varphi_0 }
\begin{cases}
K_{ \nu } ( r_0 \sigma_0 ) I_{ \nu } ( \rho \sigma_0 )
&\text{ for $ \rho \le r_0 $ }, \\
I_{ \nu } ( r_0 \sigma_0 ) K_{ \nu } ( \rho \sigma_0 )
&\text{ for $ \rho \ge r_0 $ },
\end{cases}
\end{equation}
where
$ I_{ \nu } $
and
$ K_{ \nu } $
are modified Bessel and Hankel functions, correspondingly,
$ \sigma_0 = \sqrt{ \sigma_0^2 } $,
$ \mathrm{ Re } \sqrt{ \phantom{ 0 } } > 0 $.

``Scattered'' field is the general solution of homogeneous equations in~%
\eqref{eq:Helmholtsdecomposed}
which can be easily transformed to Bessel or modified Bessel equations.
Inside the vacuum channel (area 1,
$ \rho \le a $,
$ \sigma = \sigma_0 $%
), it is convenient to decompose this general solution over linearly independent solutions of modified Bessel equation,
$ I_{ \nu } ( \rho \sigma_0 ) $
and
$ K_{ \nu } ( \rho \sigma_0 ) $.
Since it should contain no singularity for
$ \rho \to 0 $,
we obtain:
\begin{equation}
\label{eq:tildeEz1Hz1}
\tilde{ E }_{ z \omega \nu }^{ ( 1 ) }
=
A_{ \nu }^{ ( E1 ) }
I_{ \nu } ( \rho \sigma_0 ),
\quad
\tilde{ H }_{ z \omega \nu }^{ ( 1 ) }
=
A_{ \nu }^{ ( H1 ) }
I_{ \nu } ( \rho \sigma_0 ).
\end{equation}
Outside the vacuum channel (area 2,
$ \rho \ge a $%
), it is convenient to decompose this general solution over linearly independent solutions of Bessel equation,
$ H_{ \nu }^{ ( 1 ) } ( \rho s ) $
and
$ H_{ \nu }^{ ( 2 ) } ( \rho s ) $,
where
$ s = \sqrt{ s^2 } $,
$ \mathrm{ Im } \sqrt{ \phantom{ 0 } } > 0 $.
Since it should represent an outgoing wave, we obtain:
\begin{equation}
\label{eq:tildeEz1Hz1}
\tilde{ E }_{ z \omega \nu }^{ ( 2 ) }
=
A_{ \nu }^{ ( E2 ) }
H_{ \nu }^{ ( 1 ) } ( \rho s ),
\,\,\,
\tilde{ H }_{ z \omega \nu }^{ ( 2 ) }
=
A_{ \nu }^{ ( H2 ) }
H_{ \nu }^{ ( 1 ) } ( \rho s ).
\end{equation}
Boundary conditions (continuity of
$ \tilde{ E }_{ z \omega \nu } $,
$ \tilde{ H }_{ z \omega \nu } $,
$ \tilde{ E }_{ \varphi \omega \nu }$
and
$ \tilde{ H }_{ \varphi \omega \nu } $%
) for
$ \rho = a $
result in
$ 2 \times 2 $
linear system for unknown
$ A_{ \nu }^{ ( E1,2 ) } $
and
$ A_{ \nu }^{ ( H1,2 ) } $.
The determinant is:
\begin{equation}
\label{eq:Det}
\begin{aligned}
&\Delta_{ \nu }
=
\left[
\nu ( \beta a )^{ -1 }
I_{ \nu }
\left( \sigma_0^2 + s^2 \right)
\right]^2
- \\
&-
\frac{
\left[
\sigma_0^2 s \varepsilon H_{ \nu }^{ \prime } I_{ \nu } {+}
s^2 \sigma_0 I_{ \nu }^{ \prime } H_{ \nu }
\right]
\!
\left[
\sigma_0^2 s \mu H_{ \nu }^{ \prime } I_{ \nu } {+}
s^2 \sigma_0 I_{ \nu }^{ \prime } H_{ \nu }
\right]
}
{ H_{ \nu }^2 },
\end{aligned}
\end{equation}
where
\begin{equation}
\label{eq:IHprimeIH}
\begin{aligned}
I_{ \nu } &\equiv I_{ \nu }( a \sigma_0 ),
\quad
H_{ \nu } \equiv H_{ \nu }^{ ( 1 ) }( a s ),
\\
I_{ \nu }^{ \prime } &\equiv \left. \frac{ d I_{ \nu }( \xi ) } { d \xi }
\right|_{ \xi = a \sigma_0 },
\quad
H_{ \nu }^{ \prime } \equiv \left. \frac{ d H_{ \nu }^{ ( 1 ) }( \xi ) }{ d \xi }
\right|_{ \xi = a s }.
\end{aligned}
\end{equation}
Coefficients are expressed as follows:
\begin{equation}
\label{eq:AnuE1}
\begin{aligned}
&\tilde{ A }_{ \nu }^{ ( E1 ) }
{=}
\frac{ 1 }{ \Delta_{ \nu } H_{ \nu }^2 }
\left\{
\left[
\nu ( \beta a )^{ -1 } I_{ \nu } ( \sigma_0^2 + s^2 )
\right]^2
H_{ \nu }^2 K_{ \nu } I_{ \nu }^{ -1 }
\right.
{+} \\
&{+}
\left.
\left[
\sigma_0^2 s \mu H_{ \nu }^{ \prime } I_{ \nu } {+}
s^2 \sigma_0 I_{ \nu }^{ \prime } H_{ \nu }
\right]
\!\!
\left[
\sigma_0^2 s \varepsilon H_{ \nu }^{ \prime } K_{ \nu } {+}
s^2 \sigma_0 K_{ \nu }^{ \prime } H_{ \nu }
\right]
\right\},
\end{aligned}
\end{equation}
\begin{equation}
\label{eq:AnuH1}
\begin{aligned}
\tilde{ A }_{ \nu }^{ ( H1 ) }
&{=}
\frac{ \nu I_{ \nu } ( \sigma_0^2 + s^2 ) }{ i \beta a \Delta_{ \nu } H_{ \nu } }
\left\{
\left[
\sigma_0^2 s \varepsilon H_{ \nu }^{ \prime } K_{ \nu } {+}
s^2 \sigma_0 K_{ \nu }^{ \prime } H_{ \nu }
\right]
\right.
\!\!
{+} \\
&{+}
\!\!
\left.
K_{ \nu } I_{ \nu }^{ -1 }
\left[
\sigma_0^2 s \varepsilon H_{ \nu }^{ \prime } I_{ \nu } {+}
s^2 \sigma_0 I_{ \nu }^{ \prime } H_{ \nu }
\right]
\right\},
\end{aligned}
\end{equation}
where, in addition to~%
\eqref{eq:IHprimeIH},

\begin{equation}
\label{eq:KprimeK}
K_{ \nu } \equiv K_{ \nu }( a \sigma_0 ),
\quad
K_{ \nu }^{ \prime } \equiv \left. \frac{ d K_{ \nu }( \xi ) } { d \xi }
\right|_{ \xi = a \sigma_0 },
\end{equation}
\begin{equation}
\label{eq:AnuE2H2}
\tilde{ A }_{ \nu }^{ ( E2 ) }
{=}
\tilde{ A }_{ \nu }^{ ( E1 ) }
\frac{ I_{ \nu } }{ H_{ \nu } }
+
\frac{ K_{ \nu } }{ H_{ \nu } },
\quad
\tilde{ A }_{ \nu }^{ ( H2 ) }
{=}
\tilde{ A }_{ \nu }^{ ( H1 ) }
\frac{ I_{ \nu } }{ H_{ \nu } }
\end{equation}
and
\begin{equation}
\label{eq:A2tildeA}
A_{ \nu }
=
\tilde{ A }_{ \nu }
\frac{ q \omega }{ i \pi \upsilon^2 \gamma^2 }
\exp ( i \nu \varphi_0 )
I_{ \nu } ( r_0 \sigma_0 ).
\end{equation}

Using the properties of Bessel functions,
$ I_{ - \nu } = I_{ \nu } $,
$ K_{ - \nu } = K_{ \nu } $,
$ H_{ - \nu } = \exp( i \nu \pi  ) H_{ \nu } $,
one can show that
\begin{equation}
\label{eq:nu2-nu}
\begin{aligned}
\Delta_{ - \nu } &= \Delta_{ \nu },
\,
\tilde{ A }_{ - \nu }^{ ( E1 ) } = \tilde{ A }_{ \nu }^{ ( E1 ) },
\,
\tilde{ A }_{ - \nu }^{ ( H1 ) } = - \tilde{ A }_{ \nu }^{ ( H1 ) }, \\
\tilde{ A }_{ - \nu }^{ ( E2 ) } &= \frac{ \tilde{ A }_{ \nu }^{ ( E2 ) } }{ \exp( i \nu \pi ) },
\quad
\tilde{ A }_{ - \nu }^{ ( H2 ) } = \frac{ - \tilde{ A }_{ \nu }^{ ( E2 ) } }{ \exp( i \nu \pi ) }.
\end{aligned}
\end{equation}
Using~%
\eqref{eq:nu2-nu},
one can present ``scattered'' field in the bulk of medium (area 2) in the following convenient form:
\begin{equation}
\label{eq:Ez2}
\begin{aligned}
&E_{ z \omega }^{ ( 2 ) }
=
\frac{ q \omega \exp ( i \omega z / \upsilon ) }{ i \pi \upsilon^2 \gamma^2 }
\left\{
\vphantom{ \sum\limits_{ \nu = 1 }^{ \infty } }
I_0 ( r_0 \sigma_0 )
H_0^{ ( 1 ) } ( \rho s )
\tilde{ A }_0^{ ( E2 ) }
+
\right. \\
&\left.
+
2
\sum\limits_{ \nu = 1 }^{ \infty }
\cos \left[ \nu ( \varphi - \varphi_0 ) \right]
I_{ \nu } ( r_0 \sigma_0 )
\tilde{ A }_{ \nu }^{ ( E2 ) }
H_{ \nu }^{ ( 1 ) } ( \rho s )
\right\},
\end{aligned}
\end{equation}
\begin{equation}
\label{eq:Hz2}
\begin{aligned}
&H_{ z \omega }^{ ( 2 ) }
=
\frac{ q \omega \exp ( i \omega z / \upsilon ) }{ i \pi \upsilon^2 \gamma^2 }
\times \\
&\times
2
\sum\limits_{ \nu = 1 }^{ \infty }
\sin \left[ \nu ( \varphi - \varphi_0 ) \right]
I_{ \nu } ( r_0 \sigma_0 )
i \tilde{ A }_{ \nu }^{ ( H2 ) }
H_{ \nu }^{ ( 1 ) } ( \rho s ).
\end{aligned}
\end{equation}

\section{Erratum}

After the publication we have found that the paper contains sadly misprints: minus sign is missed in Eq.~%
\eqref{eq:AnuE1} 
and plus sign should be replaced by minus sign in Eq.~%
\eqref{eq:AnuH1} 
of the Appendix~%
\ref{app:etalon}.
Correct formulas are given below:
\begin{widetext}

\begin{figure*}[b]
\centering
\includegraphics[width=0.83\linewidth]{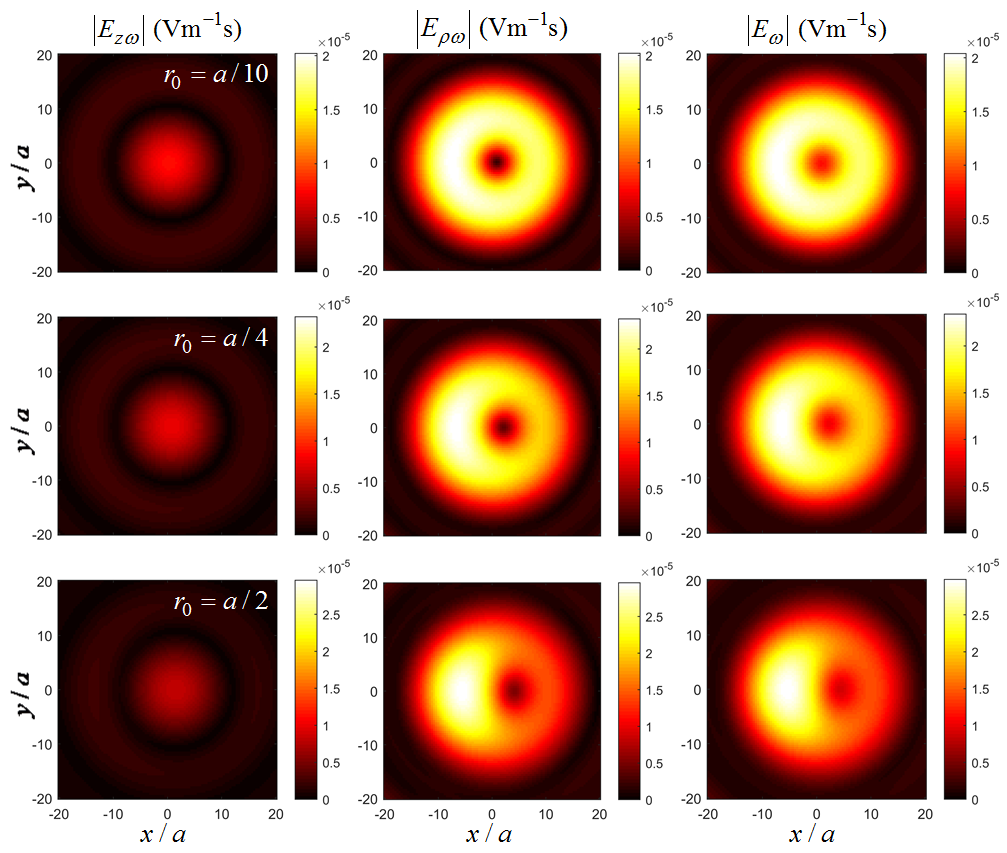}
\caption{\label{fig:FieldsAsymmErratum}%
Two dimensional distribution of absolute values of longitudinal (%
$ E_{ z \omega } $%
) and transverse (%
$ E_{ \rho \omega } $%
) and total field over
$ xy $%
-plane in the focal plane of the concentrator (%
$ z = 0 $%
) for three values of charge's shift
$ r_0 = a / 10, \, a / 4, \, a / 2 $.
Calculation parameters are the same as in Fig.~%
\ref{fig:FieldsAsymm}%
.}
\end{figure*}

$$
\label{eq:AnuE1erratum}
\tilde{ A }_{ \nu }^{ ( E1 ) }
{=}
\frac{ 1 }{ \Delta_{ \nu } H_{ \nu }^2 }
\left\{
-\left[
\nu ( \beta a )^{ -1 } I_{ \nu } ( \sigma_0^2 + s^2 )
\right]^2
H_{ \nu }^2 K_{ \nu } I_{ \nu }^{ -1 }
{+} 
\left[
\sigma_0^2 s \mu H_{ \nu }^{ \prime } I_{ \nu } {+}
s^2 \sigma_0 I_{ \nu }^{ \prime } H_{ \nu }
\right]
\left[
\sigma_0^2 s \varepsilon H_{ \nu }^{ \prime } K_{ \nu } {+}
s^2 \sigma_0 K_{ \nu }^{ \prime } H_{ \nu }
\right]
\right\},
\eqno (\mathrm{B22})
$$
$$
\label{eq:AnuH1erratum}
\tilde{ A }_{ \nu }^{ ( H1 ) }
{=}
\frac{ \nu I_{ \nu } ( \sigma_0^2 + s^2 ) }{ i \beta a \Delta_{ \nu } H_{ \nu } }
\left\{
\left[
\sigma_0^2 s \varepsilon H_{ \nu }^{ \prime } K_{ \nu } {+}
s^2 \sigma_0 K_{ \nu }^{ \prime } H_{ \nu }
\right]
{-}
K_{ \nu } I_{ \nu }^{ -1 }
\left[
\sigma_0^2 s \varepsilon H_{ \nu }^{ \prime } I_{ \nu } {+}
s^2 \sigma_0 I_{ \nu }^{ \prime } H_{ \nu }
\right]
\right\}.
\eqno (\mathrm{B23})
$$
\end{widetext}

These small corrections however influence numerical results of the paper for nonsymmetrical case presented in Fig.~%
\ref{fig:FieldsAsymm} 
and Fig.~%
\ref{fig:peaks}.
Recalculated figures (Fig.~%
\ref{fig:FieldsAsymmErratum}
and Fig.~%
\ref{fig:peaksErratum}%
, respectively)
and revised discussion are presented below. 

\begin{figure}[t]
\centering
\includegraphics[width=0.8\linewidth]{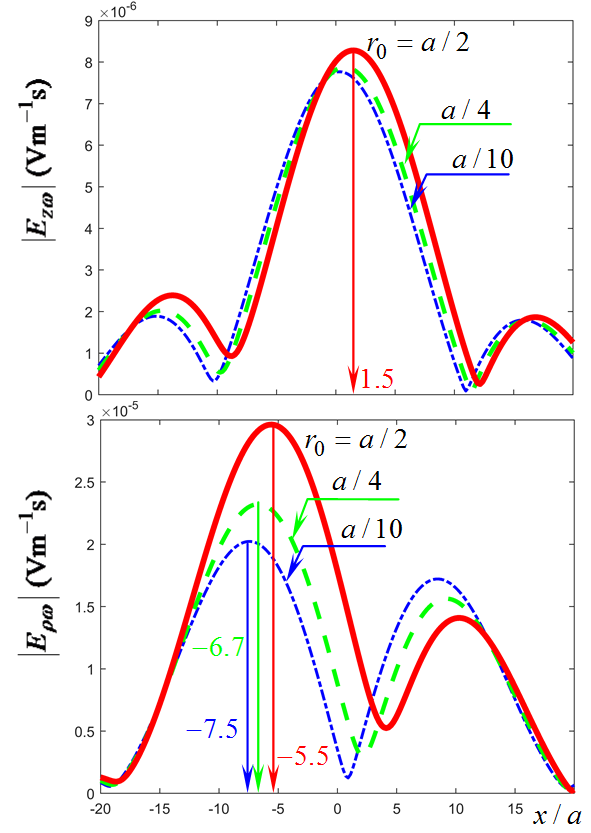}
\caption{\label{fig:peaksErratum}%
Dependence of absolute values of longitudinal (%
$ E_{ z \omega } $%
) and transverse (%
$ E_{ \rho \omega } $%
) field over
$ x $
for
$ y = z = 0 $
for three values of charge's shift
$ r_0 = a / 10, \, a / 4, \, a / 2 $.
Calculation parameters are the same as in Fig.~%
\ref{fig:peaks}%
.}
\end{figure}
%

First, asymmetry in the longitudinal field is very weak (see Fig.~%
\ref{fig:FieldsAsymmErratum}%
, left column and Fig.~%
\ref{fig:peaksErratum}%
, top plot).
The peak is slightly shifted in the area
$ x > 0 $%
, i.e. in the area where the shifted charge moves, but even for the relatively large offset (%
$ r_0 = a / 2 $%
) the peak shift is only about 
$ 1.5 a $.
The magnitude of this peak is practically not changed with 
$ r_0 $%
, thus the longitudinal field is principally the same as in the symmetrical case for the used parameters.

In turn, the transverse field (see Fig.~%
\ref{fig:FieldsAsymmErratum}%
, second column and and Fig.~%
\ref{fig:peaksErratum}%
, bottom plot) has an essential asymmetry: an expressed peak located in the region 
$ x < 0 $
can be clearly recognized especially for large offset.
The magnitude of this peak increases essentially with an increase in ratio
$ r_0 / a $.
The shift of this peak is relatively large compared to 
$ a $ 
and slightly changes with
$ r_0 $%
.
Therefore, one can conclude that asymmetry in the transverse field increases with an increase in value of charge shift.
Moreover, as one can see from Fig.~%
\ref{fig:peaksErratum}%
, separate detection of the peaks of the longitudinal field and the transverse field is problematic.
For small offsets (%
$ r_0 \lesssim a / 4 $%
) the peak of the longitudinal field is located close to the charge trajectory (i.e. for 
$ \rho < a $%
) while for larger offsets (%
$ r_0 \sim a / 2 $%
) it is dominated by strong transverse field.  

Discussion on possible applications of the presented dielectric concentrator should be also revised.
First, the longitudinal peak actually has a single peak (instead of two) therefore all statements concerning possible detection of two peaks should be canceled.
However, strong peak of the transverse field can be actually utilized for determination of beam shift and positioning of the beam toward the axis of the structure.
Moreover, this can be done for arbitrary offsets (not only small offsets as was initially stated).
The statement that ``if the bunch is well aligned along the axis'' the ``interaction between the bunch and strongly concentrated radiated field can lead to longitudinal modulation of the bunch'' can be additionally clarified.
As one can see from Figs.~%
\ref{fig:FieldsAsymmErratum}
and 
\ref{fig:peaksErratum}%
, for offsets
$ r_0 \lesssim a / 10 $
the transverse field is small comparable with the longitudinal field for 
$ \rho < a $.
Therefore mentioned modulation is potentially possible for 
$ r_0 \lesssim a / 10 $.

The authors thank S.~Baturin for his interest to the paper and fruitful discussions which have helped the authors to detect the aforementioned mistake.

%

\end{document}